# Phonon softening and superconductivity in tellurium under pressure

Francesco Mauri[a], Oleg Zakharov[a], Stefano de Gironcoli[b], Steven G. Louie[a] and Marvin L. Cohen[a]

[a] Department of Physics, University of California at Berkeley, Berkeley, CA 94720, USA
and Materials Science Division, Lawrence Berkeley Laboratory, Berkeley, CA 94720, USA

[b] Istituto Nazionale per la Fisica della Materia (INFM)
and Scuola Internazionale Superiore di Studi Avanzati (SISSA) Via Beirut 2-4, I-34014 Trieste, Italy

## Abstract

The phonon dispersion and the electron-phonon interaction for the $\beta$-Po and the bcc high pressure phases of tellurium are computed with density-functional perturbation theory. Our calculations reproduce and explain the experimentally observed pressure dependence of the superconducting critical temperature ($T_c$) and confirm the connection between the jump in $T_c$ and the structural phase transition. The phonon contribution to the free energy is shown to be responsible for the difference in the structural transition pressure observed in low and room temperature experiments.

Typeset using REVTEX

High pressure $\beta$-Po and bcc phases of tellurium (Te) are metallic and superconducting at low temperatures [1,2]. The rhombohedral $\beta$-Po structure has one atom per unit cell. The bcc structure is a special case of the $\beta$-Po structure with the angle between the lattice vectors $\alpha = 109.47°$. The transition from the $\beta$-Po to the higher pressure bcc phase is observed at 27 GPa in X-ray diffraction experiments [3] at room temperature. At the transition the unit cell volume decreases by 2% and the angle $\alpha$ increases by 4°. A jump in the superconducting critical temperature ($T_c$) from 2.5°K to 7.4°K is observed at 32-35 GPa [2]. Ref [2] suggests that the jump in $T_c$ is also related to the $\beta$-Po to the bcc phase transition and speculates that the discrepancy between the transition pressure observed in the room temperature X-ray experiments and the one found in the low temperature $T_c$ experiments is due to finite temperature effects. After the jump, $T_c$ decreases rapidly, falling below 4.5 K at 45 GPa [2]. Enhancement in $T_c$ near structural phase transitions has been observed for different systems [4], but to our knowledge no other materials exhibit such dramatic changes of $T_c$ with pressure. In this letter we present an *ab initio* calculation of phonon dispersion and electron-phonon interaction for the $\beta$-Po and the bcc phases of Te. Our calculation reproduces and explains the experimentally observed changes in $T_c$ with pressure, and confirms the connection between the jump in $T_c$ and the structural phase transition.

We use density functional theory (DFT) in the local density approximation (LDA). We compute phonon frequencies and electron-phonon coupling constants using linear-response. This approach has been successfully applied to the computation of phonon dispersions both in insulators [5] and in metals [6,7], and to the computation of the electron-phonon coupling [8]. We employ a norm conserving pseudopotential from Ref. [9]. The wavefunctions are expanded in a plane wave basis set with a cutoff energy of 20 Ry. Brillouin zone (BZ) integrations are performed using Hermite-Gaussian smearing of order one [7] with a linewidth of 0.03 Ry and 112 (408) special **k**-points in the irreducible wedge of the BZ for the bcc ($\beta$-Po) phase.

From the calculated total energies we estimate the transition pressure ($P_c$) between the $\beta$-Po and the bcc phases to be 28 GPa, in agreement with previous LDA calculations [10]. Since near the phase transition the two structures are very close, the volume-energy curves for these two structures are almost parallel [10] and a small change in the total energy for one of the phases results in a large variation of $P_c$. Thus the error in $P_c$ due to LDA could be large. Indeed $P_c$ varies by 5 GPa if the total energy of one of the two phases is shifted by 1 mRy (a typical error in LDA). The sensitivity of $P_c$ to small energy changes also suggests that the difference between the observed low temperature (32-35 GPa) and room temperature (27 GPa) values of $P_c$ can be explained by the phonon contributions to the free energy. We will explore this possibility later.

Using the approach described in Ref. [7], we compute the phonon dispersion along the high symmetry lines for the bcc and $\beta$-Po phases at different volumes. For the bcc phase we consider four unit cell volumes, 19Å$^3$, 20Å$^3$, 21Å$^3$, and 22Å$^3$. The results are presented in Fig. 1. The theoretical pressure is also shown. As the pressure decreases we observe an overall decrease of the phonon frequencies. In particular, the squares of the phonon frequencies are found to vary linearly with pressure with no peculiar behavior at $P_c$. The only nearly pressure independent mode is the $\Gamma$N transverse phonon shown by a dot-dashed line. An interesting feature of the phonon dispersion is the softening of the transverse mode (shown by the solid line) along the $\Gamma$N direction. This mode exhibits a notable phonon anomaly, i.e. a dip, in the middle of the line. As pressure decreases, the phonon frequency at the dip decreases and is imaginary at a pressure of 19 GPa. This phonon anomaly is not related directly to the structural phase transition. Indeed, both the $\beta$-Po and bcc phases have one atom per unit cell. They can not be connected with a phonon distortion with a non-zero **q** vector, since such a phonon would increase the number of atoms per unit cell. The fact that the phonon frequency becomes imaginary near the phase transition pressure is thus coincidental. Finally, at the lowest pressure another phonon anomaly along the $\Gamma$H (1$\bar{1}$1) direction is visible near the $\Gamma$ point.

It is interesting to relate the phonon dispersion to the structure of the dynamical matrix in real space, **D(R)**, **R** being a lattice vector. The squares of the phonon frequencies at a wavevector $\mathbf{q}$, $\nu^2(\mathbf{q})$, are the eigenvalues of $\mathbf{D}(\mathbf{q})/M$, $M$ being the ionic mass and $\mathbf{D}(\mathbf{q}) = -\sum_{\mathbf{R}} \mathbf{D}(\mathbf{R})[1 - \cos(\mathbf{q} \cdot \mathbf{R})]$. For simplicity we restrict our analysis to the $\Gamma$N line. If $\mathbf{q}$ is along $\Gamma$N, i.e. $\mathbf{q} = p\mathbf{G}$ where $\mathbf{G}$ is the shortest reciprocal lattice vector, and $\mathbf{R}$ is the lattice vector of a first or of a second neighbor atom, $\cos(\mathbf{q} \cdot \mathbf{R})$ is either 1 or $\cos(2\pi p)$. Thus if $\mathbf{D}(\mathbf{R})$ were short-range, i.e. $\mathbf{D}(\mathbf{R}) \neq 0$ only for first and second neighbors, $\nu^2(p)$ would just be proportional to $[1 - \cos(2\pi p)]$ [11]. The presence of a dip along $\Gamma$N indicates that the dynamical matrix also has a significant long-range contribution. In Fig. 2 we decompose $\nu^2(p)$ into a short-range "normal" part $\nu_n^2(p)$ and a long-range "anomalous" $\nu_a^2(p)$: $\nu^2(p) = \nu_n^2(p) + \nu_a^2(p)$ with $\nu_n^2(p) = A[1 - \cos(2\pi p)]$. $\nu_n^2$ is connected to the local chemistry of the crystal, whereas $\nu_a^2$ is a consequence of a $\mathbf{q}$ dependent screening induced by the electronic band structure. To fix the constant $A$ in the definition of $\nu_n^2$ we assume that at the $N$ point, i.e. the most distant point from the anomaly on the $\Gamma$N line, $\nu_a^2$ is zero. Using this decomposition, Fig. 2 shows that the anomalies of the longitudinal mode and the transverse mode (shown by the solid line) are of equal magnitude. Moreover the long-range part $\nu_a^2$ is pressure independent whereas $\nu_n^2$, exhibits a significant pressure dependence. Interestingly the anomaly in the phonon dispersion is more visible at low pressure, not because $\nu_a^2$ is larger but because $\nu_n^2$ is smaller [12].

The phonon dispersion for the $\beta$-Po structure is presented in Fig. 3 for two unit cell volumes, 21.92Å$^3$ and 23Å$^3$. The angles $\alpha$ (104.90° and 104.19° respectively) are determined by minimizing the total energy. Some similarities can be found with the phonon dispersion of the bcc phase (Fig. 1). One important difference is that for the $\beta$-Po phase a phonon anomaly is present along $\Gamma$F but not along $\Gamma$L (even though both directions are equivalent to $\Gamma$N in bcc), and that the small $\mathbf{q}$ anomaly present in bcc on the $\Gamma$H $(1\bar{1}1)$ line is absent in the $\beta$-Po on the correspondent $\Gamma$T line. The second important difference is the weaker pressure dependence of the phonon frequencies for the $\beta$-Po phase. In particular the phonon frequency at the anomaly does not vary with pressure.

Knowledge of the phonon dispersion allows us to estimate the finite-temperature free energy within the quasi-harmonic approximation [13], $F(T, V) = E(V) + F_{\text{ph}}(T, V)$, where

$E(V)$ is the DFT-LDA total energy and $F_{\text{ph}}(T,V)$ is the phonon contribution to the free energy:

$$F_{\text{ph}}(T,V) = \sum_\gamma \int \frac{d^3q}{\Omega_{\text{BZ}}} k_B T \ln\left[2\sinh\left(\frac{h\nu_{\mathbf{q},\gamma}}{2k_B T}\right)\right]. \quad (1)$$

Here $\nu_{\mathbf{q},\gamma}$ is the frequency of the phonon mode with wavevector $\mathbf{q}$ and polarization $\gamma$ computed for the unit cell volume $V$, and $\Omega_{\text{BZ}}$ is the volume of the BZ. We use the phonon dispersion computed along the high symmetry lines to estimate the BZ integral [14]. We then evaluate the transition pressure $P_c$ using $F(T,V)$ instead of $E(V)$. At T=0°K $P_c$ is 27 GPa. At room temperature (T=293°K), $P_c$ is reduced to 23 GPa. Whereas a precise determination of $P_c$ is limited by the accuracy of LDA, the change in $P_c$, $\Delta P_c = 4$ GPa, induced by the temperature is reliable. Indeed $\Delta P_c$ is not modified when the total energy of one of the two phases is shifted by 1 mRy. The computed value of $\Delta P_c$ is in good agreement with the experimental findings.

To study the superconducting properties we consider the electron-phonon coupling constants $\lambda_{\mathbf{q},\gamma}$ [15],

$$\lambda_{\mathbf{q},\gamma} = \frac{1}{\nu_{\mathbf{q},\gamma}^2} \sum_{n,m} \int \frac{d^3k}{\Omega_{\text{BZ}}} \frac{\delta(E_{\mathbf{k},n} - E_F)\delta(E_{\mathbf{k+q},m} - E_F)}{(2\pi)^2 N(E_F) M}$$
$$\times |\langle u_{\mathbf{k+q},m}|\hat{\epsilon}_{\mathbf{q},\gamma} \cdot \nabla_{\mathbf{R}} V_{sc}^{\mathbf{q}}|u_{\mathbf{k},n}\rangle|^2. \quad (2)$$

Here $\epsilon_{\mathbf{q},\gamma}$ is the polarization vector, $M$ is the atomic mass, $n$ and $m$ are the band indices, $u_{\mathbf{k},n}$ is the periodic part of the wavefunction with the eigenvalue $E_{\mathbf{k},n}$, $N(E_F)$ is the density of states per unit cell and per spin at the Fermi energy $E_F$, and $\nabla_{\mathbf{R}} V_{sc}^{\mathbf{q}}$ is the periodic part of the gradient of the self-consistent potential with respect to the atomic displacements with the wavevector $\mathbf{q}$. We compute $\lambda_{\mathbf{q},\gamma}$ along the high symmetry lines using Eq. (2), where $\nu_{\mathbf{q},\gamma}$, $\hat{\epsilon}_{\mathbf{q},\gamma}$, and $\nabla_{\mathbf{R}} V_{sc}^{\mathbf{q}}$ are obtained from the linear-response phonon calculations. We replace the $\delta$-functions with Gaussians with a standard deviations of 0.021 Ry (0.035 Ry) for the bcc ($\beta$-Po) phase. The BZ integration is performed using 5200 (2992) special $\mathbf{k}$-points in the irreducible BZ of the bcc ($\beta$-Po) structure.

In Table I we show the computed average phonon frequency ($\langle\nu\rangle$) and the integral over the BZ ($\lambda$) of the computed electron-phonon coupling constants [14]. In the bcc phase $\lambda$ increases with decreasing pressure. This increase is determined completely by the variation of the phonon frequencies. Indeed, we find that only the prefactor $1/\nu_{\mathbf{q},\gamma}^2$ in Eq. 2 changes significantly with pressure. In the $\beta$-Po phase $\lambda$ is smaller than in the bcc phase and is pressure independent. In particular, for the lowest phonon branches, which contribute the most to the value of the integrated $\lambda$, both $\lambda_{\mathbf{q},\gamma}$ and $\nu_{\mathbf{q},\gamma}$ are pressure independent. Finally in both phases $\lambda_{\mathbf{q},\gamma}$ has peaks in $\mathbf{q}$ space corresponding to the phonon anomalies, in the bcc phase along $\Gamma$N and $\Gamma$H ($1\bar{1}1$) and in the $\beta$-Po phase along $\Gamma$F. In the bcc phase the largest contribution to $\lambda$ comes from the soft phonon anomaly along $\Gamma$H. The absence of an anomaly along the corresponding $\Gamma$T direction in the $\beta$-Po phase is responsible for smaller values of $\lambda$ for this phase.

Knowledge of $\lambda_{\mathbf{q},\gamma}$ allows us to compute $T_c$ by numerically solving the Eliashberg equations [15] on a BZ mesh [14]. Our results for $T_c$ are presented in Table I and Fig. 4 for two typical values of the screened Coulomb interaction constant, $\mu^*$=0.12 and 0.14. The calculated $T_c$ is in excellent agreement with the experimental data of Ref. [2] also shown in Fig. 4.

In conclusion, using DFT-LDA and linear-response, we calculate the phonon dispersion and the electron-phonon interaction for the bcc and the $\beta$-Po high pressure phases of Te. Our results reproduce the experimental pressure dependence of the critical superconducting temperature, $T_c$, and demonstrate that the observed jump in $T_c$ corresponds to the $\beta$-Po to bcc structural phase transition. The pressure dependence of $T_c$ is shown to be primarily determined by the pressure dependence of the phonon frequencies, $\nu$. In particular: i) In $\beta$-Po the lowest phonon branches, which contribute the most to the electron-phonon coupling, are pressure independent, so is $T_c$. ii) The bcc phase has more soft phonon anomalies than $\beta$-Po. We find an enhancement of the electron-phonon interaction at the anomalies. This explains the jump in $T_c$ at the phase transition. iii) In the bcc phase, the electron-phonon coupling varies with pressure as $1/\nu^2$. The rapid increase of the short-range part of the dynamical matrix with pressure results in a rapid increase of $\nu^2$ and thus in a rapid decrease of $T_c$. Finally the difference of the transition pressures observed in low temperature and room temperature experiments is shown to arise from the phonon contribution to the free energy. The excellent agreement of our results with the experimental data illustrates the accuracy and effectiveness of DFT-LDA, linear-response approaches for study of the phonon-mediated superconductivity.

We thank Dr. V. H. Crespi for many fruitful discussions. This work was supported by the National Science Foundation (NSF) under Grant No. DMR-9120269, by the Office of Energy Research, Office of Basic Energy Sciences, Materials Sciences Division of the U.S. Department of Energy under Contract No. DE-AC03-76SF00098, by the CNR under Contract 95.01056.CT1 and by the Miller Institute for Basic Research in Science. Computer time was provided by the NSF at the National Center for Supercomputing Applications.

the values computed for the closest point along a high symmetry line.

TABLES

TABLE I. Computed average frequencies $\langle \nu \rangle$, electron-phonon coupling constants $\lambda$, and superconducting transition temperatures $T_c$ as a function of pressure. The two values for $T_c$ correspond to two different values of $\mu^*$ (0.12 and 0.14).

|  | P (GPa) | $\langle \nu \rangle$ (THz) | $\lambda$ | $T_c$ (K) |
|---|---|---|---|---|
| bcc | 47 | 3.92 | 0.93 | 6.1, 5.2 |
| bcc | 35 | 3.50 | 1.17 | 8.0, 7.1 |
| bcc | 26 | 3.10 | 1.64 | 10.0, 9.1 |
| $\beta$-Po | 23 | 3.48 | 0.80 | 3.3, 2.7 |
| $\beta$-Po | 17 | 3.28 | 0.80 | 3.4, 2.8 |

FIGURES

FIG. 1. The phonon dispersion for bcc. The directions of propagation and the polarizations are expressed in terms of the *reciprocal lattice vectors*. The dotted line denotes longitudinal modes. The solid and dot-dashed lines denote transverse modes. Along ΓN the polarization vector of the transverse mode shown by the solid line is along ($\bar{1}11$), the polarization vector of the other transverse mode is along ($0\bar{1}1$).

FIG. 2. Decomposition of $\nu^2$ into "normal" ($\nu_n^2$) and "anomalous" ($\nu_a^2$) parts for two different unit cell volumes of the bcc phase of Te along the ΓN symmetry line. The polarizations corresponding to different lines are given in the caption of Fig. 1.

FIG. 3. The phonon dispersion for the $\beta$-Po phase. The directions of propagation and the polarizations are expressed in terms of the *reciprocal lattice vectors*. Note that in the bcc lattice the directions (100) and (111) are equivalent to the directions ($1\bar{1}0$) and ($1\bar{3}1$), respectively. Along ($1\bar{1}0$) and (111), the dotted line denotes the longitudinal modes. Along (100) and ($1\bar{1}0$), the dot-dashed line denotes transverse modes; the polarization vector of the first of them is along ($0\bar{1}1$), the polarization of the second depends on **q**. Along ($1\bar{1}0$), ($1\bar{1}1$), (111), and ($1\bar{3}1$), the solid line denotes transverse modes. The other modes do not have longitudinal or transverse character.

FIG. 4. The critical superconducting temperature $T_c$ as a function of pressure. The experimental points [2] are given by solid squares with error-bars. The computed values of $T_c$ with $\mu^* = 0.12$ and $0.14$ are given by solid and open circles respectively. The lines are guides for the eye.

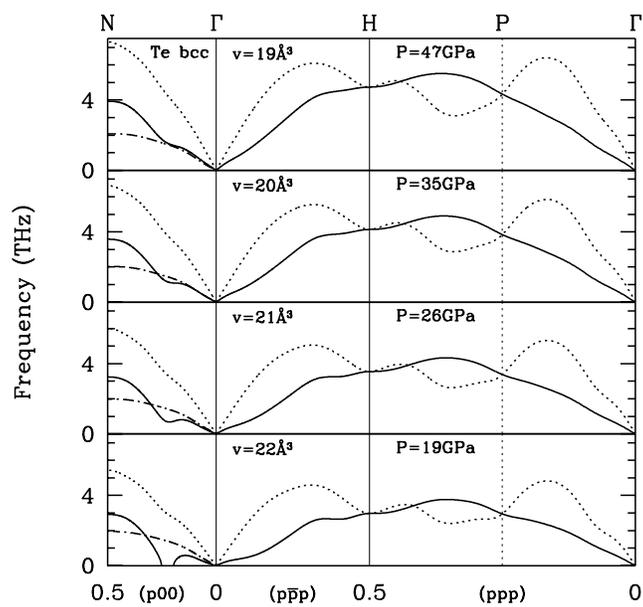

Mauri, et al.      Fig. 1

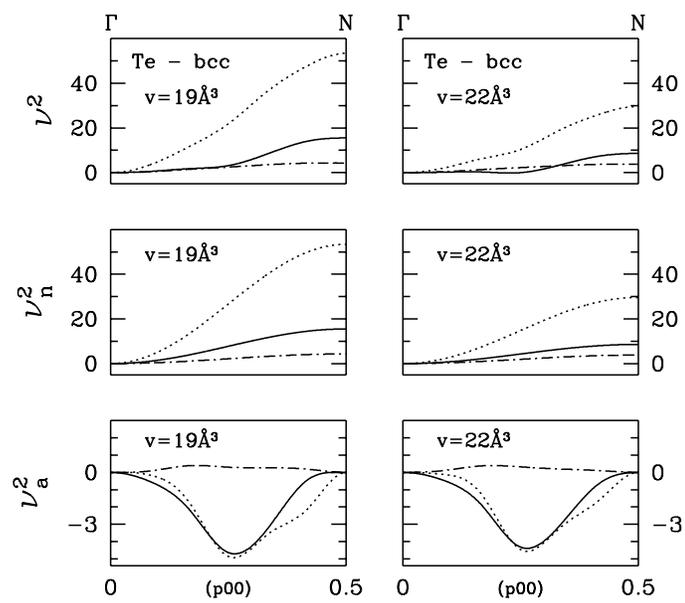

Mauri, et al.     Fig. 2

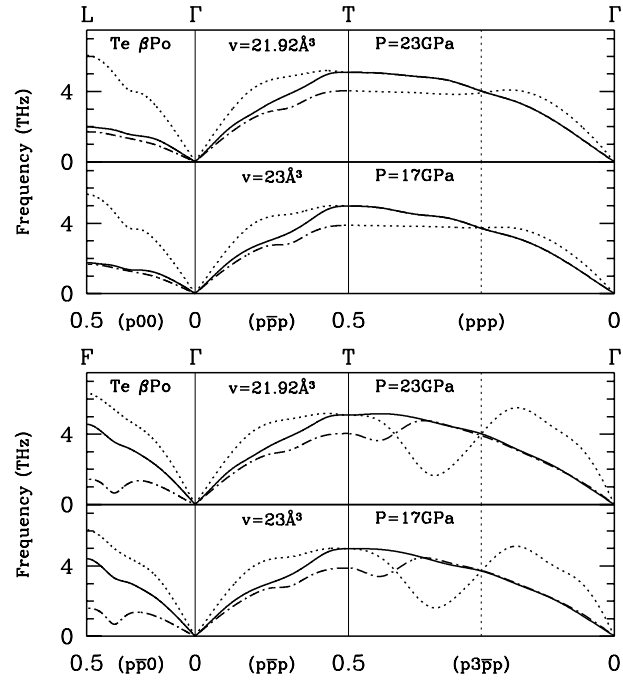

Mauri, et al.     Fig. 3

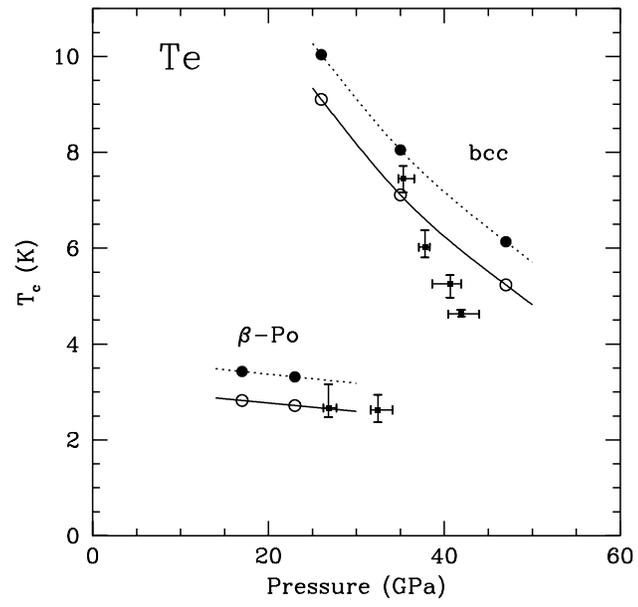

Mauri, et al.     Fig. 4